# Abaqus/Standard-based quantification of human cardiac mechanical properties

Martin Genet[1,2], LikChuan Lee[1], Ellen Kuhl[3], Julius Guccione[1]

[1] Surgery Department, School of Medicine, University of California at San Francisco, USA
[2] Marie-Curie International Outgoing Fellow
[3] Mechanical Engineering Department, Stanford University, USA

*Abstract*: *Computational modeling can provide critical insight into existing and potential new surgical procedures, medical or minimally-invasive treatments for heart failure, one of the leading causes of deaths in the world that has reached epidemic proportions. In this paper, we present our Abaqus/Standard-based pipeline to create subject-specific left ventricular models. We first review our generic left ventricular model, and then the personalization process based on magnetic resonance images. Identification of subject-specific cardiac material properties is done by coupling Abaqus/Standard to the python optimization library NL-Opt. Compared to previous studies from our group, the emphasis is here on the fully implicit solving of the model, and the two-parameter optimization of the passive cardiac material properties.*

*Keywords: Biomedical Engineering, Patient-Specific Modeling, Model Calibration, Human Left Ventricular Mechanics*

## 1. Introduction

Cardio-vascular diseaseisthe leading cause of death in the U.S., Europe, and the World. Coronary artery disease isthe most common cardio-vascular disease. It corresponds to a plaque buildup in the coronary arteries, the arteries that feed the heart muscle with fresh oxygenated blood. A heart attack occurs when an artery gets fully obstructed, for instance after plaque rupture, resulting in some severe myocardium (the heart muscle) damage or even death if the patient is not revascularized fast enough. Luckily, in the U.S. 90% of the people survive hospitalization after a heart attack. However, the damaged myocardium soon turns into fibrous tissue, which is much stiffer and much less contractile than regular myocardial tissue, leading to remodeling of the heart. And even though sometimes this remodeling is for good—like any part of the body it adapts to its new configuration,sometimes it is not, and 70% of heart failure isactually induced by heart attacks. Heart failure has recently reached an epidemic state worldwide, meaning that the prevalence for people to have heart failure increases with time.

Depending on the type of heart failure, either the heart becomes very large and poorlycontractile, orthe heart wall becomes very thick and very stiff, but the result is the same: less and less blood is pumped toward the body. Finding a treatment forheart failure is critical to improve health in the U.S., Europe, and the World.Recently, computational modeling has been shown to be of critical help in that quest, especially by providing quantitative insights into existing and potential new treatments, for instance by estimating the regional myocardial wall stress in healthy and diseased hearts(Lee, Wall, et al. 2013; Lee, Wenk, et al. 2013; Lee et al. 2014). Indeed, heart wall stress cannot be accurately measured in vivo, and more basic analytical models such as Laplace's law and the generalized Laplace's law have been shown to inaccurately predict regional heart wall stress (Zhang et al. 2011).

In this paper, we describe our Abaqus/Standard-based cardiac mechanics modeling framework. We first review our generic left ventricular (LV) model, and then our personalization procedure. The identification of passive and active cardiac material parameters is performed using Abaqus/Standard as the forward solver, and the python optimization library NL-Opt[1] as the inverse solver. In comparison to previous studies from our group, here we solve our model fully implicitly (previous versions were explicit), and we identify both the scaling and exponential terms of the Fung law.

## 2. Generic Computational Cardiac Modeling

### 2.1 Geometry, material orientation and boundary conditions

We focus here on the LV of the heart, which can be imaged in vivo and non-invasively through Magnetic Resonance Imaging (MRI). Among the four heart chambers, the LVarguably has the most important mechanical role of pumping blood to the entire body (except for the lungs) and is most susceptible to diseases and failure.Bounded at the bottom by the ventricular apex and at the top by the mitral and aortic valves, the LV has roughly the shape of a half-ellipsoid (Figure 1). We confine our analysis to the LV located just below the valves.

The myocardium has an orthotropic microstructure, where the myocytes (or myocardial cells) are arranged in stacks of two-dimensional sheets that traverse around the ventricle in a helical fashion(Humphrey 2002). The local orthogonal material basis is defined by the local fiber direction $e_F$, sheet direction $e_S$ and the sheet-normal direction $e_N$. These directional fields can be measured non-invasively using Diffusion Tensor MRI (DT-MRI) *ex vivo* hearts(Hsu et al. 1998; Holmes, Scollan, and Winslow 2000) and *in vivo*(Toussaint et al. 2013) – a significant breakthrough for patient-specific computational cardiac modeling. Despite these advancements in obtaining patient-specific myofiber orientation, the lengthy scanning time associated with DT-MRI is challenging for most patients. On the other hand, it is also possible to use previously acquired (*ex vivo* or *in vivo*) myofiber orientation maps and morph them onto another patient-specific geometry that is acquired *in vivo*(Cao et al. 2005), although this process is significantly more complicated. Due to all these reasons, we prefer to use generic rule-based myofiber orientation fields in our models(LeGrice et al. 1997), where the myofiber helix angle varies linearly from -60° at the epicardium (the external surface of the heart) to +60° at the endocardium (the internal surface of the ventricle) (Wenk et al. 2012; Lee et al. 2011) (Figure 1).

---

[1]http://ab-initio.mit.edu/nlopt

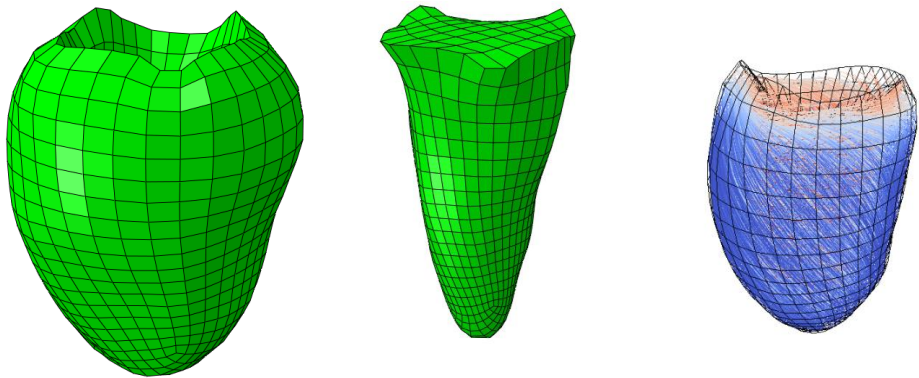

**Figure 1. Left: normal human subject-specific left ventricular geometry at beginning of diastole or filling. Middle: Hydrostatic fluid cavity mesh. Right: rule-based myofiber orientation map (color denotes helix angle, from -60°, blue, to +60°, red, with respect to the circumferential direction when viewed from the epicardium or outer surface).**

In terms of the boundary conditions imposed on our model, we fully constrain the displacement of the ventricular basal line because the mitral valve annulus and aortic valve annulus (i.e., rings that surround the valves) are much stiffer than the myocardium. The pressure load is applied to the endocardium while the epicardium is unloaded. A ventricular pressure $P$ is applied indirectly at the endocardial wall by imposing a change to the LV cavity volume $V$ at different cardiac states (as described in the next paragraph). This is performed using the hydrostatic fluid cavity modeling capabilities of Abaqus/Standard (Figure 1).

As part of the cardiovascular systemic circulation; the left ventricle fills when the left atrium contracts and the mitral valve is open and subsequently empties when the aortic valve opens and blood flows out of the LV to the body. The pressure-volume relationship is thus very complex and models have been proposed to describe the systemic circulation(Arts et al. 2005). In this study, we do not consider the full cardiac cycles but only different cardiac states in the cardiac cycle:

- Beginning of Diastole (BD) – widely considered to be the most relaxed (unloaded) state of the LV with a cavity volume $V = V_{BD}$ and pressure $P = 0$.
- End of Diastole (ED) – end of filling with an LV cavity volume $V_{ED} > V_{BD}$ and an end-diastolic pressure $P = P_{ED} > 0$.
- End of Systole (ES) – end of ejection with an LV cavity volume $V_{ES} = V_{BD}$ and an end-systolic pressure $P = P_{ES}$.

### 2.2    Passive material behavior

Two families of model have been proposed for the myocardium mechanical behavior: *(i)* strain-based Fung's law in its transversely isotropic (Guccione, McCulloch, and Waldman 1991) and fully orthotropic (Costa, Holmes, and McCulloch 2001) versions and *(ii)* invariants-based, Holzapfel-Ogden law (Holzapfel and Ogden 2009). We use the transversely isotropic Fung originally proposed in (Guccione, McCulloch, and Waldman 1991), which derives from the following free energy potential:

$$2\psi(\underline{\underline{E}}) = C_0 \left(e^{Q(\underline{\underline{E}})} - 1\right),$$

where $C_0$ is a scaling parameter that is personalized for each subject, and the exponential term $Q$ is a function of the Green-Lagrange strain $\underline{\underline{E}}$.

Several functional forms of $Q$ have been proposed. In Guccione, McCulloch, and Waldman (1991), $Q$ is defined as (with $b_1 = 0$, $b_2 = b_f$, $b_3 = b_t$ and $b_4 = b_{ft}$ in Equation (13)):

$$Q(\underline{\underline{E}}) = b_f E_{ff}^2 + b_t(E_{ss}^2 + E_{nn}^2 + E_{sn}^2 + E_{ns}^2) + b_{ft}(E_{fs}^2 + E_{sf}^2 + E_{fn}^2 + E_{nf}^2),$$

whereas $Q$ is given in (Costa et al. 1996) by (with $b_1 = b_f$, $b_2 = b_t$ and $b_3 = b_{ft}$ in Equation (2.4)):

$$Q'(\underline{\underline{E}}) = b_f E_{ff}^2 + b_t(E_{ss}^2 + E_{nn}^2 + 2E_{sn}E_{ns}) + 2b_{ft}(E_{fs}E_{sf} + E_{fn}E_{nf}).$$

These 2 forms are strictly equivalent. In Abaqus, the Fung law is implemented with:

$$Q''(\underline{\underline{E}}) = \underline{\underline{E}} : \underline{\underline{\underline{\underline{B}}}} : \underline{\underline{E}},$$

where $\underline{\underline{\underline{\underline{B}}}}$ is a fourth-order tensor. We can define the fourth-order tensor:

$$\underline{\underline{\hat{B}}} = \begin{bmatrix} b_f & 0 & 0 & 0 & 0 & 0 \\ 0 & b_t & 0 & 0 & 0 & 0 \\ 0 & 0 & b_t & 0 & 0 & 0 \\ 0 & 0 & 0 & b_{ft} & 0 & 0 \\ 0 & 0 & 0 & 0 & b_{ft} & 0 \\ 0 & 0 & 0 & 0 & 0 & b_t \end{bmatrix},$$

using the Kelvin representation of a fourth-order tensor with small symmetries (François 1995; Mehrabadi and Cowin 1990), i.e., a matrix with the independent components of the original tensor, the shear components being corrected with $\sqrt{2}$ so that $\underline{\underline{B}}$ and its matrix representation $\underline{\underline{\hat{B}}}$ are consistent. Thus, we have $b_{1111} = b_f$, $b_{2222} = b_{3333} = b_t$, $b_{1212} = b_{1313} = b_{ft}/2$ and $b_{2323} = b_t/2$, and the functional form is:

$$Q''(\underline{\underline{E}}) = b_f E_{ff}^2 + b_t \left(E_{ss}^2 + E_{nn}^2 + \frac{E_{sn}^2 + E_{ns}^2 + E_{sn}E_{ns} + E_{ns}E_{sn}}{2}\right) + b_{ft} \left(\frac{E_{fs}^2 + E_{sf}^2 + E_{fs}E_{sf} + E_{sf}E_{fs}}{2} + \frac{E_{fn}^2 + E_{nf}^2 + E_{fn}E_{nf} + E_{nf}E_{fn}}{2}\right).$$

This form is used here and is strictly equivalent to the original formulation. To simplify the personalization process, we prescribed a constant ratio between the exponential terms so that the transverse stiffness is 40% of the fiber stiffness:

$$\begin{cases} b_f = B_0 \\ b_t = 0.40\,B_0 \\ b_{ft} = \dfrac{b_f + b_t}{2} = 0.70\,B_0 \end{cases}$$

In the above equation, $B_0$ is a second independent scaling parameter that is personalized for each subject.

To account for the quasi-incompressibility of the myocardial wall (Humphrey 2002), we used the following augmented free energy potential (Holzapfel 2000; Lemaître et al. 2009)

$$\bar{\psi}\left(\underline{\underline{E}}\right) = \frac{1}{K}\left(\frac{J^2-1}{2} - \ln(J)\right) + \psi\left(\underline{\underline{\bar{E}}}\right),$$

where $K$ is the bulk modulus, $J$ is the Jacobian of the deformation gradient $\underline{\underline{F}}$ (i.e. $J = det\left(\underline{\underline{F}}\right)$) and $\underline{\underline{\bar{E}}}$ is the isochoric Green-Lagrange strain tensor that is defined as $\underline{\underline{\bar{E}}} = \left(\underline{\underline{\bar{F}}}^T \underline{\underline{\bar{F}}} - \underline{\underline{1}}\right)/2$ with $\underline{\underline{\bar{F}}} = J^{-1/3}\underline{\underline{F}}$. Note that $\underline{\underline{\bar{F}}}$ is the isochoric component of the deformation gradient $\underline{\underline{F}}$ such that $det\left(\underline{\underline{\bar{F}}}\right) = 1$. In all our models, we prescribed $K = 10^{-3}$, a value close to the bulk modulus of water.

The Fung law is implemented as a user material subroutine in C++ using the LMT++ library for basic linear algebra (Leclerc 2010; Genet 2010). We validated our user material subroutine using the original Abaqus implementation of the Fung law. Figure 2 shows a comparison of the stress-strain responses between our implementation and Abaqus implementation of the Fung's law in a single hexahedral element successively loaded in the six canonical loadings (three pure extensions, three pure shears).

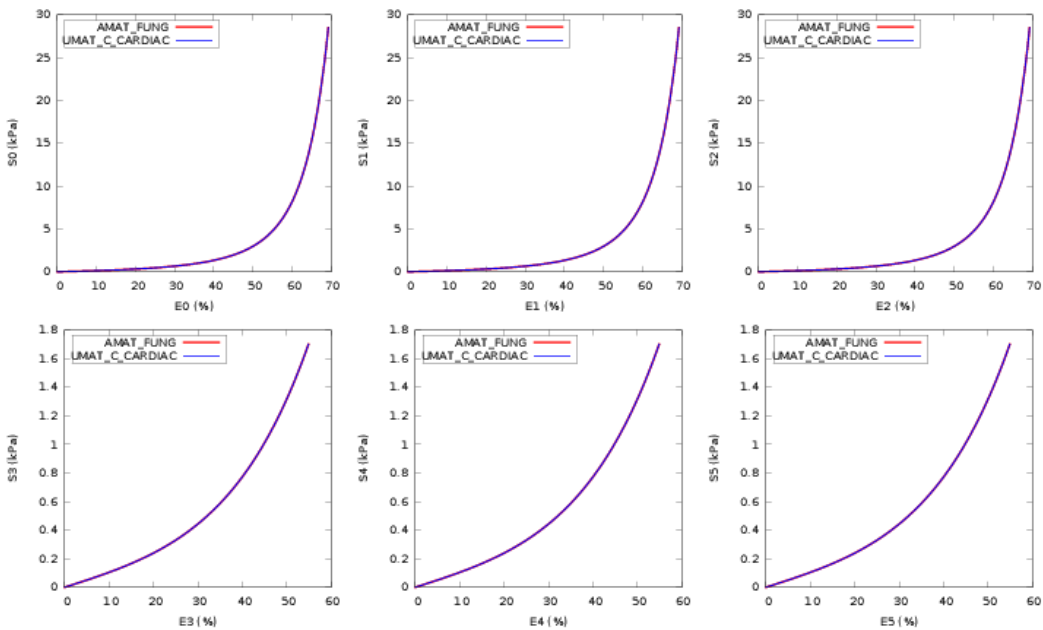

**Figure 2. Validation of the implementation of the transversely isotropic Fung law as a user material subroutine, based on Abaqus internal implementation.**

## 2.3    Active material behavior

There are several approaches to model the active contraction of cardiac tissues. The most rigorous approach is, perhaps,to model the fully coupled electromechanical behavior of the cardiac tissue. In this approach, one solves a coupled system of partial differential equations that describe the diffusion of the action (or electric) potential, development of the active force and balance of mechanical momentum (Göktepe and Kuhl 2009; Wall et al. 2012). In the simplified approach, all the myocytes are assumed to be activated instantaneously and thediffusion of the action potentialthen becomes irrelevant and can be neglected. We used the simplified approach here.

Many models have been proposed to describe the process of active force generation. These models are largely phenomenological and range from simple to very complex (Rice and de Tombe 2004). We use the model formulated by (Guccione and McCulloch 1993; Guccione, Waldman, and McCulloch 1993). In this model, the active stress $T$ depends on the current sarcomere length calculated through the fiber strain $E_{ff}$, peak intracellular concentration $Ca_0$, a reference tension $T_{max}$ and the time $t$ from which the tissue is activated through the following equation

$$T(t, E_{ff}) = \frac{T_{max}}{2} \frac{Ca_0^2}{Ca_0^2 + ECa_{50}^2(E_{ff})} \left(1 - \cos\left(\omega(t, E_{ff})\right)\right),$$

where

$$ECa_{50}(E_{ff}) = \frac{Ca_{0_{max}}}{\sqrt{e^{B(l(E_{ff})-l_0)} - 1}}$$

$$\omega(t, E_{ff}) = \begin{cases} \pi \dfrac{t}{t_0} & \text{when } 0 \leq t \leq t_0 \\ \pi \dfrac{t - t_0 + t_r\left(l(E_{ff})\right)}{t_r} & \text{when } t_0 \leq t \leq t_0 + t_r\left(l(E_{ff})\right) \\ 0 & \text{when } t \geq t_0 + t_r\left(l(E_{ff})\right) \end{cases}$$

$$t_r(l) = ml + b$$

$$l(E_{ff}) = l_r\sqrt{2E_{ff} + 1}$$

The reference tension $T_{max}$ is personalized for each subject. Following (Walker et al. 2005), we assume that an active force with a magnitude that is 40% of that in the fiber direction also develops in the sheet direction.

The law was implemented in the user material subroutineand validated against the analytical law on a single isometric contraction virtual experiment as shown inFigure 3.

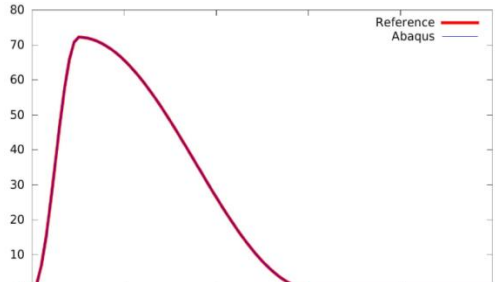

**Figure 3. Validation of the implementation of the active force generation law as a user material subroutine, based on the analytical solution.**

## 3. Model Personalization

### 3.1 From MRI to FEM

The first step in personalizing the previously described generic computational LV model is to reconstruct the relaxed or unloaded geometry from MRI data. This is done by manually segmenting the LV endocardium and epicardium inboth short-axis and long-axis image slices using the medical image analysis software MeVisLab[2]. Thereafter, the segmented contours were usedto create a fully hexahedral finite element mesh of the subject's relaxed LV geometry with the meshing software TrueGrid[3]. Figure 4showstheresults of a convergence study on the mesh density based on the LV model of one subject. The figure shows that a good balance between numerical accuracy and computational cost can be achieve in a mesh containing 5000 nodes. This mesh density was used for all our models.

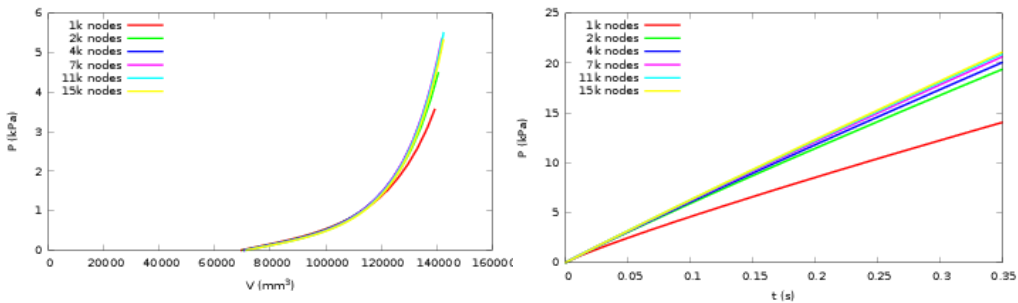

**Figure 4. Sensitivity analysis of the ventricular passive (left) and active (right) responses to the mesh density.**

---

[2] http://www.mevislab.de
[3] http://truegrid.com

## 3.2 Passive material behavior personalization

The second step of the personalization process is to calibrate the passive material law to a given subject. This is done by optimizing the material parameters to get the MRI-measured ED volumeand the prescribed ED pressure of 9 mmHg.

One important point is that the optimization process is more stable when the ventricle is loaded by prescribing the LV cavity volume as opposed to prescribing the LV cavity pressure. This is because the material deformation is constrained by the prescribed LV cavity volume even when the material parameters vary during the optimization process. Thus, the elements are less likely to be distorted and the solver is less likely to diverge during the optimization process. In contrast, the elements can become highly distorted in the case when pressure is prescribed (instead of volume), especially when the material parameters have small values that make the LV become un-physiologically soft.

Unlike previous studies where only $C_0$ was calibrated and the exponential terms ($b_f$, $b_t$ and $b_{ft}$) were defined based on previous large animal studies (Sun et al. 2009; Lee et al. 2011; Wenk et al. 2012), we propose to calibrated both $C_0$ and $B_0$. However, there is an infinite combination of these 2 parameters for each given ED volume and ED pressure. To find a unique and subject-specific value for these 2 parameters, we also minimize the distance of the end-diastolic pressure-volume response to the Klotz curve (Klotz, Hay, and Dickstein 2006). Indeed, it was showed that when scaled properly, the LVof different species behave similarly (Klotz, Hay, and Dickstein 2006).

The optimization contains two intertwined loops.The first, outer,loop is used to optimize $B_0$ so that the distance to the Klotz curve is minimized. The second, inner,loop is used to optimize $C_0$ for a given $B_0$ so that the calculated ED pressure matches the prescribed one. The optimization process is illustrated by the $C_0 - B_0$ map inFigure 5. In Fig. 5, the red line represents the $(C_0, B_0)$ combination that matches the prescribed ED pressure. The optimization process consists of searching along the red line fora point at which the distance between the calculated end-diastolic pressure-volume response and the Klotz curve is minimal. An example Klotz curve distance map is given Figure 5.

The optimization loops are constructed using the NL-Opt python library, where Abaqus/Standard plays the role of the forward solver.

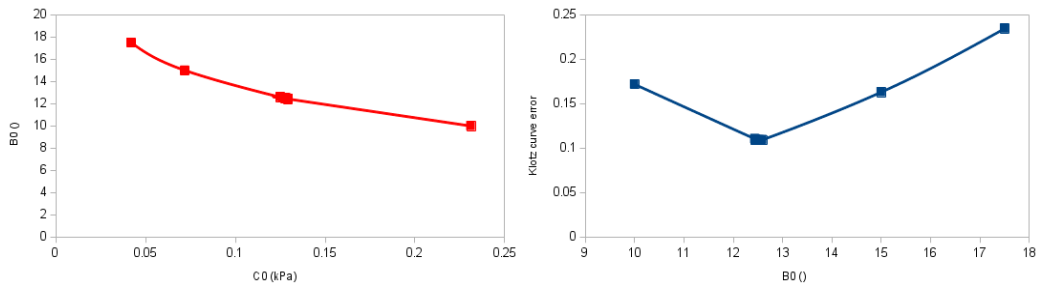

**Figure 5. Left: $C_0 - B_0$ couples that verify the experimental ED volume-pressure conditions. Right: Along that line, distance to the Klotz curve as function of $B_0$.**

### 3.3 Active material behavior personalization

The last personalization step is to calibrate the active material law. It is simpler than the passive material law since there is only one parameter $T_{max}$ to optimize. The parameter $T_{max}$ is optimized so as to match ES volume and ES pressure. ES volume is assumed to be equal to BD volume, and ES pressure is assumed to be 120 mmHg. The computation is again driven by volume in order to limit the range of deformation reached throughout the optimization process.

## 4. Applications

The personalization procedure described in this paper was applied to five normal human subjects. The LV finite element meshes are represented Figure 6. Table 1 provides, for each subject as well as the mean and standard deviation values, the MRI-extracted ES and ED volumes, and corresponding ejection fractions.

Table 2 provides, for each subject as well as the mean and standard deviation values, the identified passive and active cardiac material parameters. The variability between these five normal subjects is quite low.

Note that the validation of these personalized models based on regional strain measured through tagged MRI has been studied in (Genet et al.). Basically, it was found that the average difference between the measured and predicted circumferential strains is below 5%.

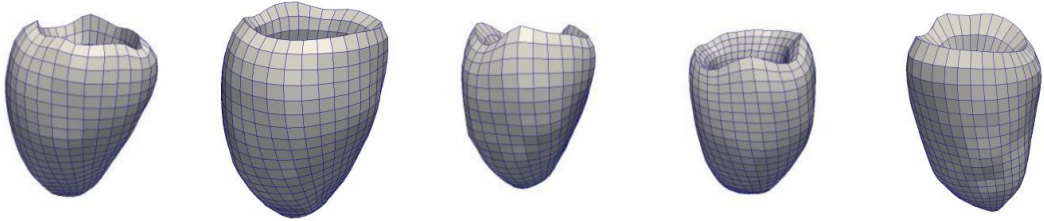

**Figure 6. Personalized finite-element meshes geometries of five normal human subjects.**

**Table 1. End-disatolic and End-systolic volumes measured though MRI, as well as associated ejection fraction, for the five normal human subjects.**

|  | #1 | #2 | #3 | #4 | #5 | Mean | Standard deviation |
|---|---|---|---|---|---|---|---|
| ES volume (ml) | 63 | 51 | 65 | 40 | 40 | 52 | 12 |
| ED volume (ml) | 146 | 114 | 134 | 86 | 105 | 117 | 24 |
| EF (%) | 57 | 55 | 51 | 53 | 62 | 56 | 4 |

**Table 2. Personalized passive and active cardiac material parameters for the five normal human subjects.**

|  | #1 | #2 | #3 | #4 | #5 | Mean | Standard deviation |
|---|---|---|---|---|---|---|---|
| $C_0$ (kPa) | 0.117 | 0.121 | 0.105 | 0.107 | 0.123 | 0.115 | 0.00817 |
| $B_0$ () | 12.4 | 15.0 | 18.3 | 16.0 | 10.1 | 14.4 | 3.18 |
| $T_{max}$ (kPa) | 130 | 149 | 148 | 132 | 155 | 143 | 11.1 |

## 5. Conclusion

In this paper, we described our pipeline to build subject-specific computational left ventricular models from magnetic resonance images. One important point is that we found a very low variability in the three material parameter values that define passive and active heart muscle mechanical properties. The main implications are that these material parameters could be used to compute normal human left ventricular stress, which could be used as a target for *in silico* optimization of cardio-vascular disease treatment. Moreover, these material parameters could be used for both normal human subjects and different cardio-vascular disease states to create a database to guide clinicians in the diagnosis and treatment of patients.

In conclusion, Abaqus/Standard can be used to efficiently model cardiac mechanics. Moreover, coupled to the open-source python optimization library NL-Opt, it can be used to reliably personalize computational cardiac models based on MRI. This opens the door for many personalized, quantitative, *in silico* studies of cardiac procedures.